\documentclass[twocolumn,showpacs,unsortedaddress,superscriptaddress,showkeys,nofootinbib,preprintnumbers,letterpaper]{revtex4-1}
\usepackage{bm}
\usepackage{upgreek}
\usepackage{acronym}
\usepackage{amsmath}
\usepackage{amssymb}
\usepackage{graphics}
\usepackage{subfigure}
\usepackage{here}
\usepackage[squaren]{SIunits}
\usepackage[usenames]{color}
\usepackage{url}
\usepackage{xspace}
\usepackage[normalem]{ulem}
\usepackage[%
 setpagesize=false,%
 bookmarks=true,%
 bookmarksdepth=tocdepth,%
 bookmarksnumbered=true,%
 colorlinks=false,%
 pdftitle={},%
 pdfsubject={},%
 pdfauthor={},%
 pdfkeywords={}%
]{hyperref}

\newcommand{\wCO}{\texttt{w/ CO}\xspace}
\newcommand{\woCO}{\texttt{w/o CO}\xspace}

\newcommand*{\aye}{\mathrm{i}}
\newcommand*{\diff}{\,\mathrm{d}}

\newcommand{\bmuptheta}{\bm{\uptheta}}

\newcommand{\figref}[1]{Fig.\ \ref{#1}}
\newcommand{\secref}[1]{Sec.\ \ref{#1}}

\addunit{\annum}{a}
\addunit{\AU}{AU}
\addunit{\parsec}{pc}

\begin{document}
\preprint{RESCEU-23/23}


\title{On the Testability of the Quark-Hadron Transition\\
Using Gravitational Waves From Merging Binary Neutron Stars}

\author{Reiko Harada}
\email{harada.reiko@resceu.s.u-tokyo.ac.jp}
\affiliation{Research Center for the Early Universe (RESCEU), Graduate School of Science, The University of Tokyo, Tokyo 113-0033, Japan}
\affiliation{Department of Physics, Graduate School of Science, The University of Tokyo, Tokyo 113-0033, Japan}
\author{Kipp Cannon}
\email{kipp@resceu.s.u-tokyo.ac.jp}
\affiliation{Research Center for the Early Universe (RESCEU), Graduate School of Science, The University of Tokyo, Tokyo 113-0033, Japan}
\author{Kenta Hotokezaka}
\email{kentah@resceu.s.u-tokyo.ac.jp}
\affiliation{Research Center for the Early Universe (RESCEU), Graduate School of Science, The University of Tokyo, Tokyo 113-0033, Japan}
\author{Koutarou Kyutoku}
\email{kyutoku@tap.scphys.kyoto-u.ac.jp}
\affiliation{Department of Physics, Kyoto University, Kyoto 606-8502, JAPAN}
\affiliation{Center for Gravitational Physics and Quantum Information, Yukawa Institute for Theoretical Physics, Kyoto University, Kyoto 606-8502, JAPAN}
\affiliation{Interdisciplinary Theoretical and Mathematical Sciences Program (iTHEMS), RIKEN, Saitama 351-0198, JAPAN}

\begin{abstract}
	Elementary particles such as quarks and gluons are expected to be fundamental degrees of freedom at ultra high temperatures or densities, while natural phenomena in our daily lives are described in terms of hadronic degrees of freedom.
	Massive neutron stars and remnants of binary neutron star mergers may contain quark matter, but it is not known how the transition from hadron matter to quark matter occurs.
	Different transition scenarios predict different gravitational waveforms emitted from binary neutron star mergers.
	If the difference between the equations of state occurs at sufficiently high density, it is expected that the difference between waveforms mainly appears in the merger or the post-merger phase rather than in the inspiral phase.
	The typical frequency of gravitational waves after the coalescence is higher than \unit{2}{\kilo\hertz}, which is difficult to observe using current detectors.
	In this study, we performed Bayesian model selection for two representative scenarios and investigated whether observations with future detectors will allow us to identify the correct model.
	We assume that the relatively low density equation of state around the nuclear saturation density is completely known from accumulated observations.
	Under this assumption, we find that it is reasonable to expect to be able to identify the correct transition scenario with third-generation detectors or specialized detectors with high sensitivity at high frequencies designed for post-merger signal observation, \textit{e.g.}, NEMO.

\end{abstract}


\maketitle
\acrodef{SNR}{signal-to-noise-ratio}

\section{Introduction}\label{sec:introduction}

Neutron stars (NSs) have masses similar to that of the Sun within radii of about $\unit{10}{\kilo\meter}$.
They are the densest observable object in the universe and provide unique laboratories for testing the properties of cold and dense matter.
Observed massive NSs have masses $\sim 2 \, M_\odot$ \cite{Demorest_2010, Antoniadis_2013, Fonseca_2021}, and their inferred core densities are $\gtrsim (3\ \text{to}\ 4) n_0$ \cite{Lattimer_2005}, where $n_0\sim\unit{0.16}{\femto\meter}^{-3}$ is the nuclear saturation density.
The nature of the transition from nuclear matter (NM) to quark matter (QM) is not yet well understood.
However, in such an extremely dense environment, quark degrees of freedom may be released beyond pure hadronic matter \cite{Baym_2018}.

The phase evolution of gravitational waves in the last few orbits of merging binary neutron stars (BNSs) is affected by the stellar distortion caused by the tidal force of the companion.
Its measurement provides a way to infer the tidal deformability of NSs \cite{Flanagan_2008, Chatziioannou_2020, Damour_2012, Read_2013, Del_Pozzo_2013, Bernuzzi_2015, Hotokezaka_2016, Hinderer_2016, Lackey_2017}.
Since these macroscopic properties of stars are determined by the equation of state (EoS) of the matter, we can constrain the EoS.
There are a number of previous studies on possible constraints given by the tidal deformability measurements on the nature of the transition from NM to QM \cite{Christian_2019, Han_2019, Chatziioannou_2020_phase, Pang_2020, Raithel_2023, Pereira_2022, essick2023phase} (see also \cite{Wijngaarden_2022} for a study combining pre- and post-merger signals).
It has been shown that a strong first-order phase transition that has relatively large effects on the property of typical mass NSs can be verified by about 10 events detected by current detectors with a signal-to-noise ratio (SNR) above 30 \cite{Pang_2020}.
In similar optimistic cases, it is suggested that hybrid objects with sharp phase transitions and mixed states may start being distinguished by the gravitational wave data with tidal deformability uncertainty smaller than 5\%--10\% \cite{Pereira_2022}.

However, because tidal effects on massive NSs are small and difficult to measure accurately \cite{essick2023phase, huxford2023accuracy}, they are not very informative to examine the properties of matter at the ultra high densities which are only realized in the core of a massive neutron star $\gtrsim 2\,M_\odot$ \cite{Annala_2020}.
It has also been claimed that such heavy stars are rare in double neutron star systems \cite{Lattimer_2012}.
On the other hand, the fate of the remnant of the BNS merger has been observed in numerical relatively (NR) simulations to depend sensitively on the high-density EoS \cite{Sekiguchi_2011, Radice_2017, Bauswein_2019, Blacker_2020, blacker2023exploring, Most_2019, Weih_2020, Prakash_2021, espino2023revealing, Huang_2022, PhysRevLett.130.091404, Vijayan_2023}.
Depending on the stiffness of the EoS, it will collapse quickly and become a black hole, or it will live a long time as a massive NS \cite{2011PhRvD..83l4008H}.
Therefore, it has been proposed that gravitational waves from post-merger remnants can be used to probe the matter at ultra high densities \cite{Breschi_2019, easter2021measure, breschi2022kilohertz, dhani2023prospects}.
A previous investigation examined the ability of the Einstein Telescope to distinguish example models with and without hyperon, and suggests that observation of a single merger remnant at a distance of up to $\sim \unit{200}{\mega\parsec}$ can rule out one of the two posibilities with strong evidence \cite{Radice_2017}.

In \cite{PhysRevLett.130.091404}, the EoSs for two representative scenarios of the transition from NM to QM were constructed taking into account \textit{ab initio} constraints of the chiral effective theory ($\chi$EFT) \cite{Drischler_2021} and perturbative quantum chromodynamics (pQCD) \cite{Gorda_2021}, and gravitational waveforms for these two scenarios were obtained with NR simulations.
One scenario is a smooth crossover and the other is a strong first-order phase transition.
It should be noted that these are just representative classifications that are expected to be relatively easy to distinguish by observations.
The category ``crossover'' here does not exclude a second-order nor a weak first-order phase transition.
The second candidate, ``strong first-order phase transition'', here has a large discontinuous jump in density.
Such a large jump is expected to occur only at densities so high that they cannot be achieved in NS cores, because otherwise it would be difficult to explain the massive NSs observed so far (see \cite{PhysRevLett.130.091404} for a detailed discussion).
Then, strictly speaking, the NR simulations mentioned above were performed only for the EoSs with and without crossover from the NM branch provided by $\chi$EFT to the QM branch provided by pQCD.
We call these scenarios \wCO and \woCO respectively in this paper.
The nuclear EoS at low densities is identical in the two scenarios, the only difference being the presence of a smooth transition into the pQCD branch.
The simulation results suggest that the nature of the transition from NM to QM could be inferred by the time to collapse into a black hole.
However, whether the inference can be made in a realistic observational framework has not, yet, been rigoursly investigated.
The goal of this work is to see whether the representative EoSs in \cite{PhysRevLett.130.091404} can be distinguished by observation of gravitational waves emitted from BNS mergers.

\begin{table*}
	\caption{Mass configurations for numerical relativity simulations used in this work.
	$m_1$ is the primary mass of the binary and $m_2$ is the secondary mass.
	The models with $(m_1,m_2) = (1.375, 1.375)\,M_\odot$ and $(1.55,1.2)\,M_\odot$ were presented in \cite{PhysRevLett.130.091404}.	}
	\label{table:masses}
	{\tabcolsep=0.015\linewidth
	\begin{tabular}{lcccccccccccc}
		\hline
		$m_1\,[M_\odot]$& 1.25 & 1.3 & 1.35 & 1.375 & 1.4 & 1.45 & 1.5 & 1.55 & 1.4 & 1.45 & 1.5 & 1.6\\
		$m_2\,[M_\odot]$& 1.25 & 1.3 & 1.35 & 1.375 & 1.35 & 1.3 & 1.25 & 1.2 & 1.4 & 1.45 & 1.5 & 1.6\\
		$m_1+m_2\,[M_\odot]$& 2.5 & 2.6 & 2.7 & 2.75 & 2.75 & 2.75 & 2.75 & 2.75 & 2.8 & 2.9 & 3 & 3.2\\
		$m_2/m_1$& 1 & 1 & 1 & 1 & 0.964 & 0.897 & 0.833 & 0.774 & 1 & 1 & 1 & 1\\
		\hline
	\end{tabular}
	}
\end{table*}

\begin{figure*}
	\resizebox{.49\linewidth}{!}{\includegraphics{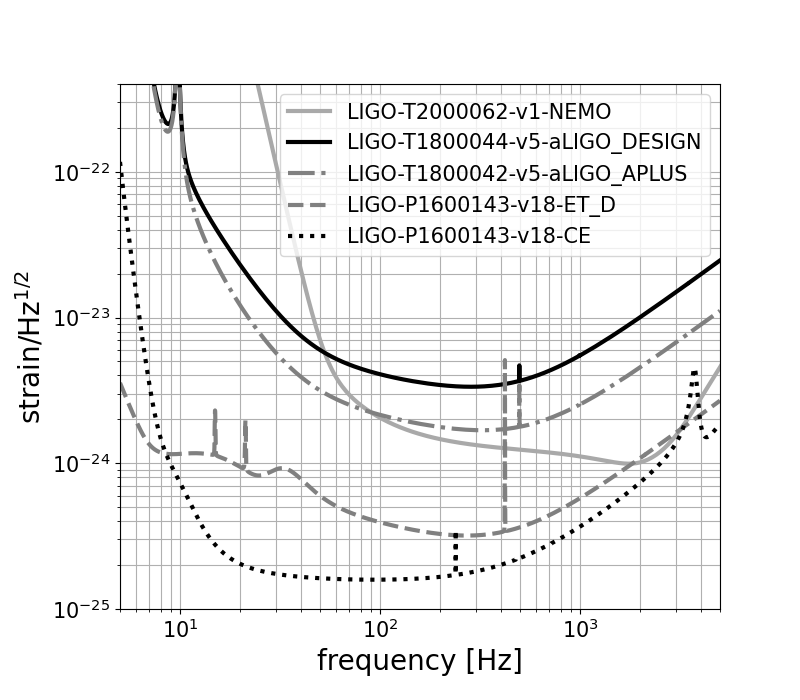}}
	\resizebox{.49\linewidth}{!}{\includegraphics{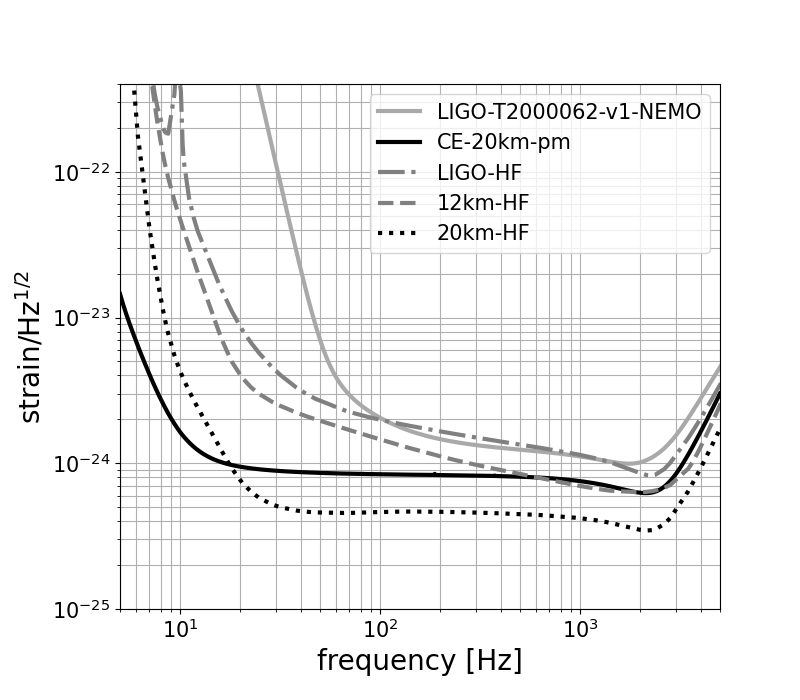}}
	\caption{The strain equivalent noise curves used in this work.
	Neutron Star Extreme Matter Observatory (NEMO) \cite{Ackley_2020, LIGO-T2000062} is shown in both panels for comparison.
	Left panel:  Advanced Laser Interferometer Gravitational-Wave Observatory (aLIGO) \cite{LIGO-T1800044}, Advanced LIGO Plus (A+) \cite{LIGO-T1800042}, Einstein Telescope (ET\_D)\cite{Hild_2011, LIGO-P1600143} and \unit{40}{\kilo\meter} long Cosmic Explorer (CE) \cite{Abbott_2017, LIGO-P1600143}.
	Right panel:  \unit{20}{\kilo\meter} CE optimized for post-merger oscillations (CE-20km-pm) \cite{Srivastava_2022}, LIGO-HF, 12km-HF and 20km-HF \cite{Martynov_2019}.}
	\label{fig:curves}
\end{figure*}

As shown in the spectrograms in \cite{PhysRevLett.130.091404}, the typical frequency of the signals emitted by the remnant of a BNS merger is higher than \unit{2}{\kilo\hertz} and it is difficult to observe with current detectors.
Even GW170817, the event with the highest SNR at the time of its detection, had no signature of the post-merger signal \cite{GW170817_observation, GW170817_postmerger}.
Therefore, in this work, we investigate the testability of the quark-hadron transition with future gravitational wave detectors.
Some promising candidates are the third generation detectors such as Einstein Telescope \cite{Hild_2011} and Cosmic Explorer \cite{Abbott_2017, Srivastava_2022}.
These detectors are more sensitive than current detectors over a wide frequency range.
Furthermore, there are several proposed detectors designed for BNS post-merger signals and they have high sensitivity at high frequencies \cite{Martynov_2019, Ackley_2020, Srivastava_2022}.
The sensitivity curves of the detector designs used in this work are shown in \figref{fig:curves} together with the corresponding LIGO document numbers.

We construct a hypothesis test and from it derive an indicator measuring how distinguishable \wCO is from \woCO.  We measure the distribution of this indicator for each detector and each gravitational wave simulation (see \secref{subsec:criterion}).
We find that the value of this indicator is inversely proportional to the distance to the source, \textit{i.e.}, proportional to the SNR (see \secref{subsec:dl}).
From that, we make an optimistic estimate of the expected number of events per year corresponding to each value of the indicator based on the BNS merger rate $(10\,\text{--}\,1700)\,\mathrm{Gpc^{-3}a^{-1}}$ estimated by the LIGO-Virgo-KAGRA (LVK) Collaborations \cite{PhysRevX.13.011048} (\figref{fig:rate} in \secref{subsec:dl}).
Furthermore, we extremize the indicator's expectation value by reducing the degrees of freedom of the hypothesis test, but we find this effort does not help much.

This paper is organized as follows.
\secref{sec:method} provides information about the NR waveforms used in this work and methods for evaluating distinguishability and testing usefulness of the dimensionality reduction.
\secref{sec:results} presents the results of performed tests and their interpretation.
\secref{sec:conclusion} gives a summary of the whole paper.

\section{Method}\label{sec:method}
\subsection{Numerical Relativity Simulations}
\begin{figure*}
	\resizebox{1.\linewidth}{!}{\includegraphics{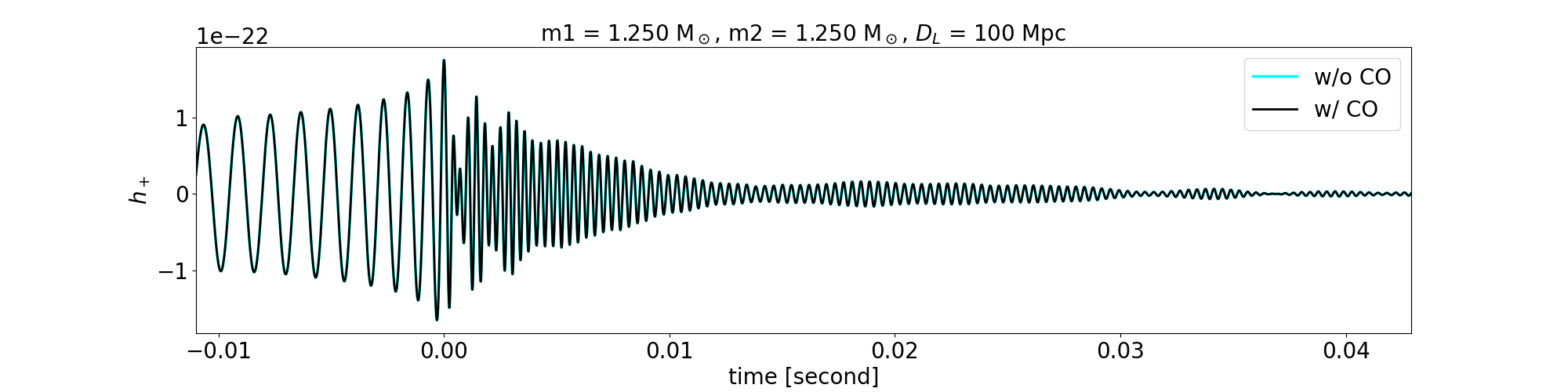}}
	\resizebox{1.\linewidth}{!}{\includegraphics{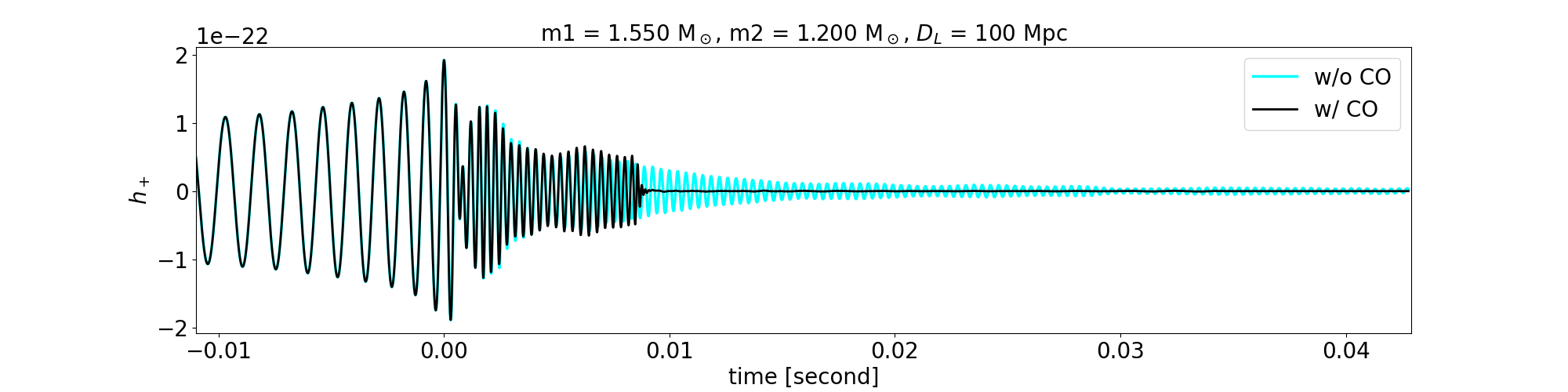}}
	\resizebox{1.\linewidth}{!}{\includegraphics{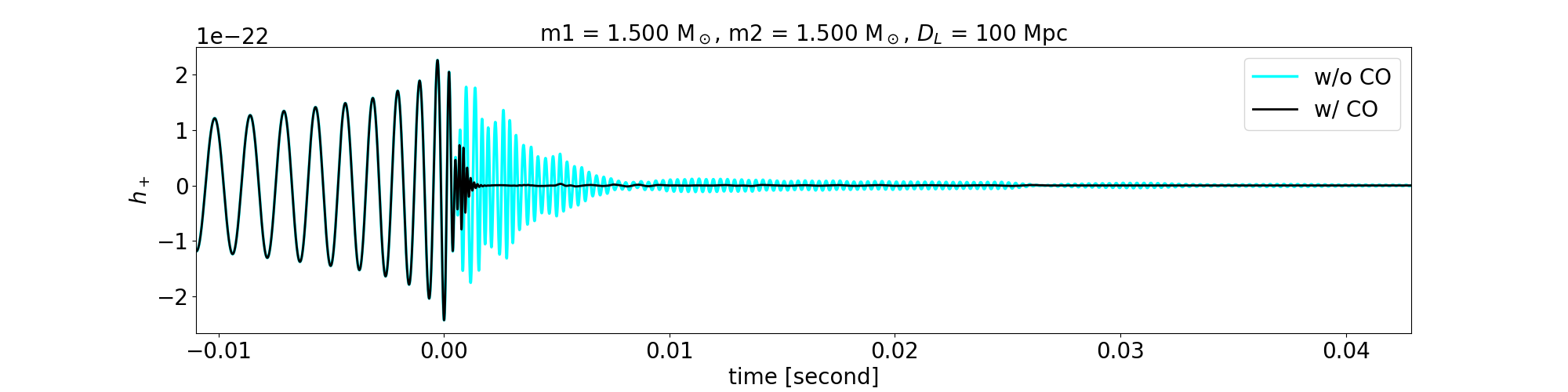}}
	\resizebox{1.\linewidth}{!}{\includegraphics{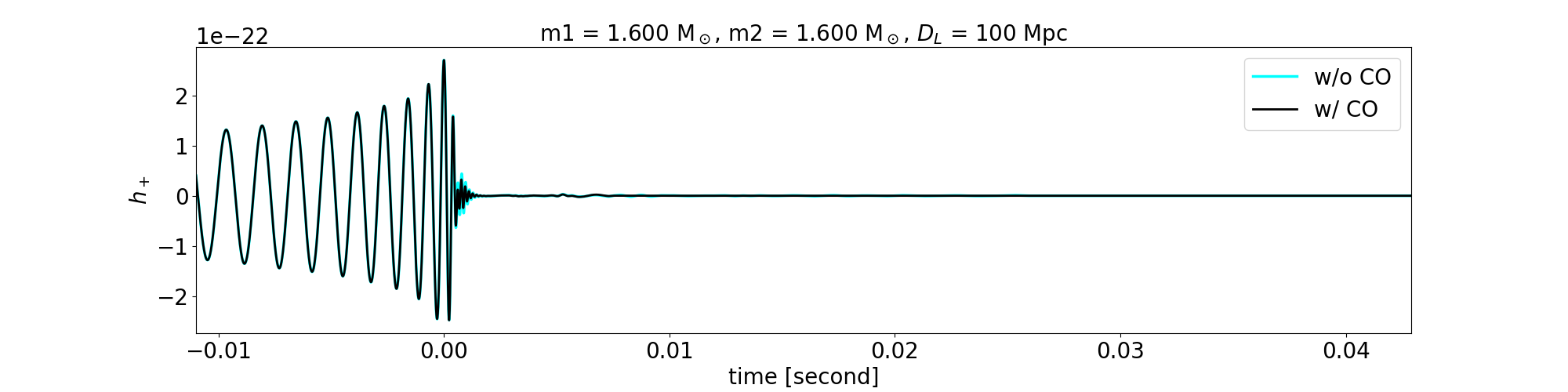}}
	\caption{NR waveforms for masses of $(m_1, \,m_2)=(1.25, 1.25)\,M_\odot$, $(1.55, 1.2)\,M_\odot$, $(1.5, 1.5)\,M_\odot$ and $(1.6, 1.6)\,M_\odot$ and the luminosity distance of $D_L=\unit{100}{\mega\parsec}$.}
	\label{fig:waveform}
\end{figure*}

In this work, we used the gravitational waveforms obtained as results of NR simulations from several cycles before coalescence to post-merger.
The method of simulations is described in \cite{PhysRevLett.130.091404}'s supplemental material.
To explore the dependence on the total mass and the mass ratio, we simulated ten new configurations in addition to the two presented in \cite{PhysRevLett.130.091404}.
They are summarized in Table \ref{table:masses}.
The details of these new simulations will be presented elsewhere.
Some example waveforms are shown in \figref{fig:waveform}.

\subsection{Criterion for Evaluating Distinguishability}\label{subsec:criterion}
This section describes the criterion to evaluate the distinguishability of the two models, \wCO and \woCO, in this work.
We performed the following analyses:
\begin{enumerate}
	\item Generate fake data $\mathbf{d}$ by injecting the NR waveform of the model \wCO into stationary Gaussian noise having the spectrum of the detector in question,
		\begin{equation}
			\mathbf{d}= \text{Gaussian noise}+ \text{\wCO NR waveform}. \label{eq:data}
		\end{equation}
		The NR waveforms are sampled at discrete points.
		When a function of time $X(t)$ is sampled at discrete points, we will write it as a finite-dimensional vector $\mathbf{X}$.
		\label{proc:data}
        \item Calculate the log Bayes factors,
        \begin{eqnarray}
            \log B = \log \frac{Z_1}{Z_2}, &
		Z_{s} \equiv\int \diff\bmuptheta L(\mathbf{d}|\bmuptheta, \mathcal{H}_s)\pi(\bmuptheta),\label{eq:B}
        \end{eqnarray}
        where $L$ is the likelihood function, $\bmuptheta$ are the waveform parameters, $\mathcal{H}_s$ is the hypothesis that some signal $s \in \{1,2\}$ is present in the data.
		$Z_1$ and $Z_2$ are the Bayesian evidences $Z$ for \wCO and \woCO.

		    We assume a Dirac delta function prior $\pi(\bmuptheta) = \delta(\bmuptheta-\bmuptheta_0)$, where $\bmuptheta_0$ are the true values of $\bmuptheta$.
		    We will justify this choice later in this section.

		The range of frequencies used in the calculation was \unit{10}{\hertz} to \unit{5000}{\hertz}.
		    \label{proc:model}
        \item Repeat procedure \ref{proc:data} and \ref{proc:model} above many times for the same parameter $\bmuptheta_0$ and calculate statistics such as mean and variance of $\log B$.\label{proc:repeat}
\end{enumerate}

Because we assumed Gaussian noise, 
\begin{equation}
	L(\mathbf{d}|\bmuptheta, \mathcal{H}_s) \propto \exp\left\{-\frac{1}{2}(\mathbf{d}-\mathbf{h}(\bmuptheta)|\mathbf{d}-\mathbf{h}(\bmuptheta))\right\},
\end{equation}
where $(\mathbf{X}|\mathbf{Y})$ represents the matched filter inner product \cite{Maggiorevol1}.
When functons of time $X(t)$ and $Y(t)$ are sampled at $N$ discrete points with sampling interval $\Delta t$, their Fourier components are written as 
\begin{eqnarray}
	\tilde{X}_k&=&\sum^{N-1}_{j=0} X(t_j)\exp\left(-\aye\frac{2\pi jk}{N}\right)\Delta t,\\
	\tilde{Y}_k&=&\sum^{N-1}_{j=0} Y(t_j)\exp\left(-\aye\frac{2\pi jk}{N}\right)\Delta t
\end{eqnarray}
for $k = -\lfloor N/2 \rfloor, ..., -1, 0, 1, ...,\lfloor N/2 \rfloor$, and their matched filter inner product is
\begin{equation}
	(\mathbf{X}|\mathbf{Y}) \equiv 2 \sum_k \Delta f \frac{\tilde{X}^*_k\tilde{Y}_k}{S_k},\label{eq:matchedfilter}
\end{equation}
where $\Delta f\equiv 1/(N\Delta t)$.
$S_k = 2\langle | \tilde{d}_{k} |^{2}\rangle\Delta f$ (when $\mathbf{d}$ is only noise) is the spectral density of the noise.
The notation $\langle\cdot\rangle$ denotes the ensemble average over noise realizations.
The quantity $(\mathbf{d}-\mathbf{h}(\bmuptheta)|\mathbf{d}-\mathbf{h}(\bmuptheta))$ can be interpreted as the square magnitude of the distance between the data $\mathbf{d}$ and the waveform model $\mathbf{h}$ in the waveform space normalized by the magnitude of noise.
When $\log B > 0,\label{eq:condition1}$ the data support \wCO, rather than \woCO, and vice versa.

\begin{figure}
	\resizebox{1.\linewidth}{!}{\includegraphics{"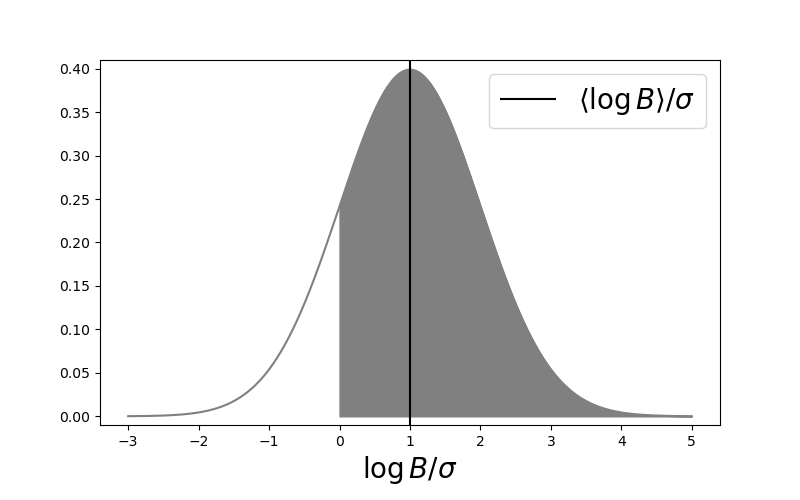"}}
	\caption{The expected distribution of $\log B$ when $\langle \log B\rangle = \sigma$.}
	\label{fig:gaussexam}
\end{figure}

Because the Bayes factor is a random variable in the presence of the detector noise, we need to calculate the distribution of $\log B$.
When assuming stationary Gaussian noise and the existence of the (known) signal in the data $\mathbf{d}$, the components of the vector $\tilde{x}_k\equiv(\tilde{d}_k-\tilde{h}_k)/\sqrt{S_k}$ are expected to be Gaussian distributed.
Then the quantity $\sum_{k} \tilde{x}_k^2=(\mathbf{d-h|d-h})$ is expected to be $\chi^2$ distributed.
Assuming further that the dimension $N$ of $\tilde{\mathbf{x}}$ is very large, the distribution of $\mathbf{(d-h|d-h)}$ should be nearly Gaussian.
As a result, we expect $\log B$ to be Gaussian distributed, and this is confirmed empirically in \secref{subsubsec:distribution}.
The mean and standard deviation are written as $\langle \log B\rangle$ and $\sigma$ below.

Since the data have been generated by \eqref{eq:data}, we know that the correct model is always \wCO, therefore $\langle \log B\rangle > 0$.
Considering the case in which $\langle \log B\rangle/\sigma=1$, the distribution of $\log B$ is shown in \figref{fig:gaussexam}.
The gray shaded area in \figref{fig:gaussexam}, $\log B/\sigma\geq0$, occupies 84\% of the total area under the curve:  when 100 events corresponding to the same parameter $\bmuptheta_0$ (masses, spins, distance to the source, etc.) are observed, 84 of them are expected to support \wCO.
If $\langle \log B\rangle/\sigma \geq 2$, over 97\% of the observed events for the same parameters will favor \wCO.
Therefore, a large value of $\langle\log B\rangle/\sigma$ means a high degree of distinguishability.

In this work, we adopt $\langle \log B\rangle/\sigma \geq 1$ as the minimum criterion for saying that a given detector can distinguish the two models \wCO and \woCO, for a given parameter set $\bmuptheta_0$.

Finally, we briefly discuss the validity of the assumption that $\pi(\bmuptheta) = \delta(\bmuptheta-\bmuptheta_0)$ we stated above.
In the limit of large SNR, the parameter estimation error can be estimated by \cite{Maggiorevol1}
\begin{equation}
	\langle \Delta\theta^i\Delta\theta^j\rangle=(\Gamma^{-1})^{ij},
\end{equation}
where
\begin{equation}
	\Gamma_{ij}\equiv(\partial_i \mathbf{h}|\partial_j \mathbf{h})\propto \mathrm{SNR}^2 \label{eq:Fisher}
\end{equation}
is the Fisher information matrix.
The nominal SNR of a signal $\mathbf{h}$ is \cite{Maggiorevol1, creighton2011gravitational}
\begin{equation}
	\mathrm{SNR}\equiv\sqrt{(\mathbf{h|h})}\label{eq:defSNR}
\end{equation}
using the inner product of  \eqref{eq:matchedfilter}.
As we will see later, for an event from which one can expect to be able to distinguish these two models, the SNR will be quite high, \textit{e.g.}, an SNR of around 400 will be needed when observed with ET\_D.
With such high SNR, parameters such as masses can be measured with high precision.
In the case of GW170817, with an SNR of 32 the inferred chirp mass could already be measured to nearly 4 significant figures $M_c\equiv (m_1 m_2)^{3/5}/(m_1 + m_2)^{1/5}=1.188^{+0.004}_{-0.002}\,M_\odot$, while the mass ratio could be constrained to $q \equiv m_2/m_1=0.7\,\text{--}\,1.0$ in the low-spin prior case, and $m_1/m_2=0.4\,\text{--}\,1.0$ in the high-spin prior case (90\% credible intervals) \cite{GW170817_observation}.
With an SNR $13\times$ higher the uncertainties in these paramters will be reduced more than an order of magnitude, thus the assumption $\pi(\bmuptheta)=\delta(\bmuptheta-\bmuptheta_0)$ is reasonable to some extent.
In addition, the number of NR simulations is limited.

\subsection{Estimation of the Event Rate}

The data $\mathbf{d}$ are the sum of the noise $\mathbf{n}$ and the real gravitational wave signal $\mathbf{h}$.
Without loss of generality, waveform model 1 can be written as $\mathbf{h}_1=\mathbf{h}+\delta \mathbf{h}_1$ and waveform model 2 can be written as $\mathbf{h}_2=\mathbf{h}+\delta \mathbf{h}_2$.
$\mathbf{h,\, h}_1$ and $\mathbf{h}_2$ are all inversely proportional to the luminosity distance $D_L$.
We further assume zero-mean stationary noise $\langle \mathbf{n}\rangle = \mathbf{0}$.
Then,
\begin{align}
\log B  &= \log \frac{L_1}{L_2}\\
	&\propto -\left[ (\mathbf{d-h}_1|\mathbf{d-h}_1)-(\mathbf{d-h}_2|\mathbf{d-h}_2)\right]\\
	&= -\left[ (\mathbf{n-\delta h}_1|\mathbf{n-\delta h}_1) -(\mathbf{n-\delta h}_2|\mathbf{n-\delta h}_2)\right]\\
	&= -\left[ (\mathbf{n|n})-2(\mathbf{n|\delta h}_1)+(\delta \mathbf{h}_1|\delta \mathbf{h}_1)\right] \nonumber\\&\quad\quad +\left[ (\mathbf{n|n})-2(\mathbf{n|\delta h}_2)+(\delta \mathbf{h}_2|\delta \mathbf{h}_2)\right]\\
	&= 2(\mathbf{n|\delta h}_1)-(\delta \mathbf{h}_1|\delta \mathbf{h}_1) -2(\mathbf{n|\delta h}_2)+(\delta \mathbf{h}_2|\delta \mathbf{h}_2)
\end{align}
Because $\langle (\mathbf{n|\delta h}_i)\rangle=0$,
\begin{equation}
	\langle \log B\rangle \propto -(\delta \mathbf{h}_1|\delta \mathbf{h}_1)+(\delta \mathbf{h}_2|\delta \mathbf{h}_2)\propto D_L^{-2}.
	\label{eq:logBaverage_DL}
\end{equation}
Furthermore,
\begin{align}
\sigma^2 &\equiv \left\langle(\log B-\langle \log B \rangle)^2\right\rangle \\
	&\propto 4\left\langle \left[(\mathbf{n|\delta h}_1) - (\mathbf{n|\delta h}_2)\right]^2\right\rangle \\
	&= 4\left\langle (\mathbf{n}|\delta\mathbf{h}_1-\delta\mathbf{h}_2)^2\right\rangle\label{eq:n_h1-h2} \\
	&= 2(\delta\mathbf{h}_1-\delta\mathbf{h}_2|\delta\mathbf{h}_1-\delta\mathbf{h}_2)\label{eq:h1-h2} \\
	&\propto D_L^{-2}.
	\label{eq:sigma_DL}
\end{align}
From \eqref{eq:n_h1-h2} to \eqref{eq:h1-h2}, the definition of PSD \cite{Maggiorevol1} was used.
Then 
\begin{equation}
	\frac{\langle\log B\rangle}{\sigma}\propto \frac{1}{D_L}\label{eq:indicator_DL}.
\end{equation}
Particularly, since $\delta \mathbf{h}_1=\mathbf{0}$ in our analysis described in \secref{subsec:criterion},
\begin{equation}
	\sigma = \sqrt{2\langle \log B \rangle}.
\end{equation}
The proportionality coefficient of \eqref{eq:indicator_DL} depends on the detector and the component masses of the binary.
In practice, it also depends on the inclination of the orbit and the sky location, but they are fixed here.
We calculated $\langle \log B \rangle/\sigma$ for various $D_L$ assuming optimal orientation for the ``+'' polarization and determined the proportionality coefficient by the least squares method.

Given values of $m_1$, $m_2$ and $\langle \log B \rangle/\sigma$, there is a corresponding luminosity distance $D_L(m_1,\,m_2,\, \langle \log B \rangle/\sigma)$.
By using it, the number of events per year with $\langle \log B \rangle/\sigma>a$ can be written as
\begin{equation}
	\mathcal{R}_{\mathrm{BNS}}\times \frac{4\pi}{3}D_{L}^{3}(m_1,\,m_2,\, a).
\end{equation}
The merger rate of BNS reported in \cite{PhysRevX.13.011048} is $\mathcal{R}_\mathrm{BNS}= (10\,\text{--}\,1700)\,\mathrm{Gpc^{-3}a^{-1}}$.
As we describe later, the assumptions of the optimal orientation and the Dirac delta-function prior make this estimation optimistic.


\subsection{Dimensionality Reduction}\label{subsec:dimensionmethod}

Considering the space of (all possible) waveforms, the waveforms predicted by the \wCO and \woCO hypotheses will differ from each other in some way, by some difference vector.  This difference vector cannot be an arbitrary function of time, because it is constrained by the physics of the two models, \textit{i.e.}, it must reside within some subspace of the space of waveforms.  There must be, therefore, directions in waveform space that are orthogonal to the subspace in which the difference vector is confined --- components of the waveforms that the two hypotheses predict to be the same.  Because the two hypotheses predict waveforms that do not differ from each other in those directions, for the purpose of the hypothesis test those waveform components are uninformative.  These parts contribute to $(\mathbf{d-h|d-h})$ only by adding random numbers.
We may be able to obtain smaller standard deviation $\sigma$ by reducing the dimensions $N$ because, as mentioned in \secref{subsec:criterion}, the quantity $(\mathbf{d-h|d-h})$ is $\chi^2$ distributed and its standard deviation is $\sqrt{2N}$.
If we can reduce $\sigma$ by reducing the dimensions, then $\langle \log B\rangle/\sigma$ can be increased, increasing the detectability of the EoS effects.

Ideally a singular-value decomposition would be performed to identify the most useful basis in which to conduct the hypthesis test but because of the limited number of NR simulations such an analysis cannot be performed at this time.  Nevertheless, visual inspection of the waveforms and an understanding of the physical processes can give us some clues.
Because the EoS at low densities is the same for both \wCO and \woCO (as mentioned in \secref{sec:introduction}), the early part of the waveform must be identical, which can be confirmed in \figref{fig:waveform}.
Likewise, there is some of time after the collision when the GW emission predicted by both models has decayed to nearly 0.
Time samples, therefore, would seem to be a good basis with which to truncate the models:  project the problem into a lower dimensional space by elliminating the uninformative early and late portions of the waveforms.  It only remains to be seen which time samples are best to remove, and how much the hypothesis test is improved by doing so.

Unfortunately we did not observe a significant improvement with this technique, but the procedure and the results are shown below.

We calculated the Bayes factor 1,000 times for each interval $[t_\mathrm{start}, t_\mathrm{end}]$, varying the starting time $t_\mathrm{start}$ and the ending time $t_\mathrm{end}$, and searched for the interval giving the maximum value of $\langle \log B\rangle/\sigma$.
Because boundary effects caused by the abrupt start of the waveform due to the finite separation of initial data too close to what proved to be the optimum $t_\mathrm{start}$ were found to introduce confusion in the extremization, we extended the waveforms to earlier times by constructing a hybrid waveform combining the analytic waveform IMRPhenomPv2\_NRTidal \cite{Dietrich_2017} generated by LALSimulation \cite{lalsuite} with the NR simulation.
The analytic waveform we added is the same for the two models.
Therefore, the contribution these additional data make to $(\mathbf{d-h|d-h})$ cannot bias the hypothesis test towards either model.
We set $t=0$ to be time  of peak amplitude and searched between $t=\unit{-0.1}{\second}$ and the end of NR simulation.

\section{Results and Discussions}\label{sec:results}
\subsection{Distribution of $\log B$}\label{subsubsec:distribution}
\begin{figure}[t]
	\resizebox{1.\linewidth}{!}{\includegraphics{"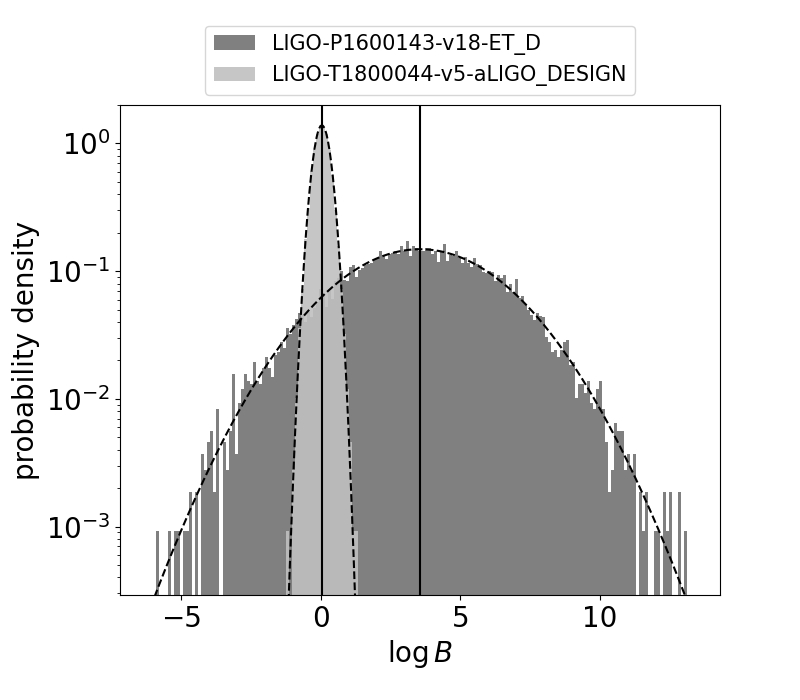"}}
	\caption{The normalized histogram of $\log B$ for Advanced LIGO and ET\_D when assuming optimal orientation for the ``+'' polarization, $(m_1,\,m_2)=(1.55,\,1.2)\,M_\odot$ and $D_L = \unit{100}{\mega}{\parsec}$.
	The number of different noise realizations is 10,000.
	The solid black lines indicate the means of the respective histograms, and the dashed black lines Gaussians whose means and variances agree with the respective histograms.
	Both distributions were found to be approximately Gaussian as expected.
	Also, for ET\_D, the distribution more strongly favours $\log B > 0$, showing it is better able to distinguish the two models in this case.}
	\label{fig:hist}
\end{figure}
To verify the assumption of Gaussianity for $\log B$, we injected gravitational waveforms from the optimal orientation for the ``+'' polarization from an $(m_1,\,m_2)=(1.55,\,1.2)\,M_\odot$ simulation at a luminosity distance $D_L=\unit{100}{\mega\parsec}$ into 10,000 realizations of noise for aLIGO\_DESIGN and ET\_D (The nominal SNRs defined by \eqref{eq:defSNR} for each case are 31 and 409 respectively).  The result is shown in \figref{fig:hist}.
For this analysis, we used almost the entire NR waveform shown in the \figref{fig:waveform}.

Both distributions were found to be approximately Gaussian as expected.
It can be seen that for ET\_D, the outcome is much more frequently greater than 0 than for aLIGO\_DESIGN.
For ET\_D, $\langle \log B \rangle/\sigma = 1.3$ and the percentage of $\log B>0$ is 91\%.
For aLIGO\_DESIGN, $\langle\log B\rangle/\sigma =0.1$ and the percentage of $\log B>0$ is 56\%.
For aLIGO\_DESIGN, it is more difficult to distinguish the two models.

\subsection{Luminosity Distance \textit{vs.}\ Distinguishability}\label{subsec:dl}
\begin{figure}
\resizebox{\linewidth}{!}{\includegraphics{"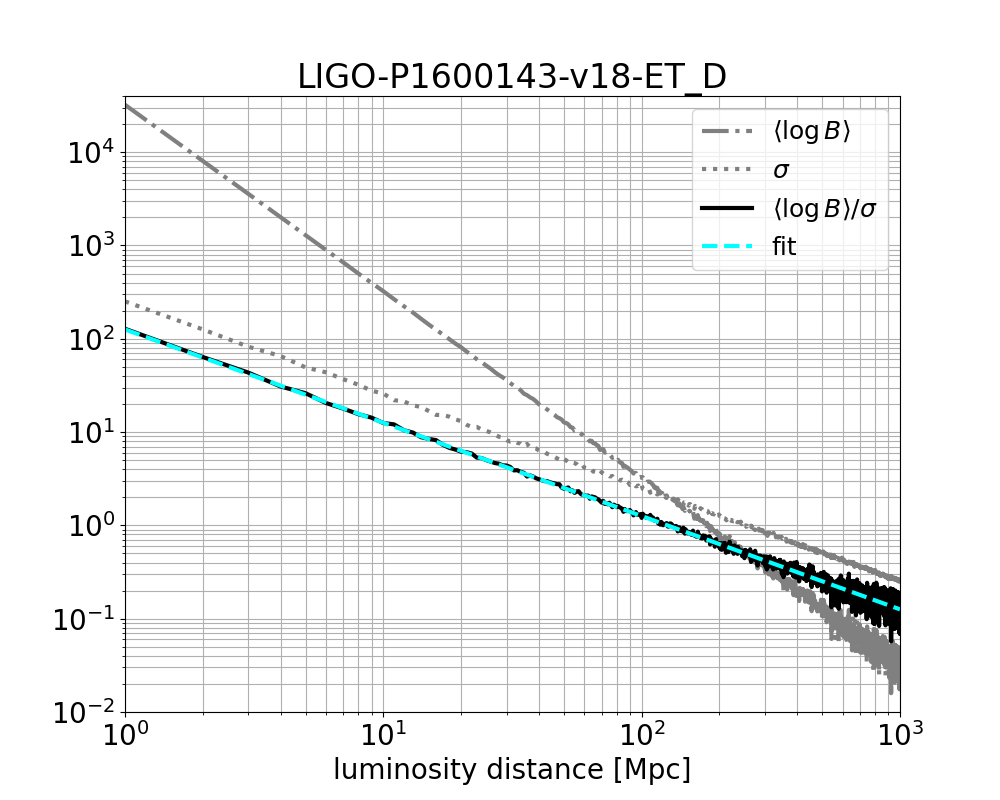"}}
\caption{The dependence on the luminosity distance $D_L$ of $\langle \log B \rangle$, $\sigma$ and $\langle \log B \rangle/\sigma$ when assuming ET\_D, the optimal orientation for the ``+'' polarization, $(m_1,\,m_2)=(1.55,\,1.2)\,M_\odot$.
The cyan line shows the results of the least squares fitting.
The relations of \eqref{eq:logBaverage_DL}, \eqref{eq:sigma_DL} and \eqref{eq:indicator_DL} can be confirmed.}
\label{fig:distance_155_12_ET_D}
\end{figure}

\figref{fig:distance_155_12_ET_D} demonstrates the relationship between distinguishability and the luminosity distance in \eqref{eq:indicator_DL}.
For this example, the optimally oriented ``$+$'' polarization of the $(m_1,\,m_2)=(1.55,\,1.2)\,M_\odot$ simulation was added to 1000 realizations of ET\_D noise over a variety of distances and the parameters shown in the legend computed.
The figure shows the result of the fit used to determine the proportionality coefficient in \eqref{eq:indicator_DL}.
The relations in \eqref{eq:logBaverage_DL} and \eqref{eq:sigma_DL} are also exhibited.

The proportionality coefficient in \eqref{eq:indicator_DL} depends on the waveform and the detector.  The numerical analysis depicted in \figref{fig:distance_155_12_ET_D} was repeated for every combination available to obtain these unknown coefficients.

\subsubsection{Dependence on Total Mass}

\begin{figure}
\resizebox{\linewidth}{!}{\includegraphics{"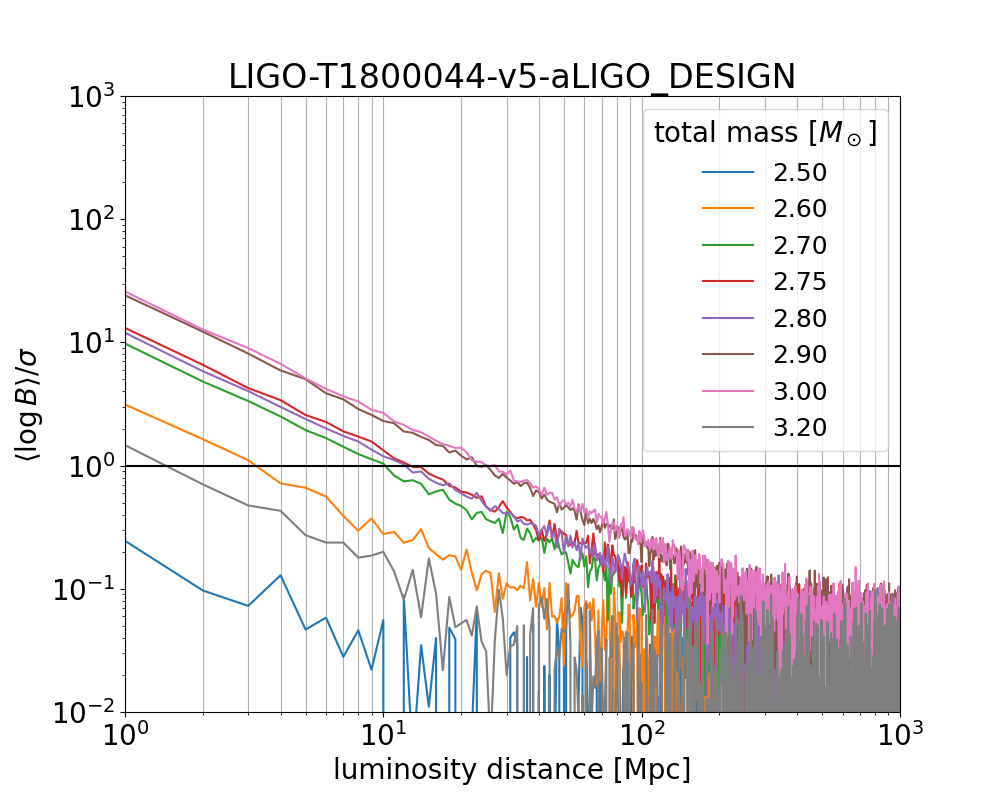"}}
\caption{$D_L$ \textit{vs.}\ $\langle \log B \rangle/\sigma$ plot for LIGO-T1800044-v5-aLIGO\_DESIGN and various total masses assuming optimal orientation for ``+'' polarization.
All binaries plotted here are equal-mass systems.
	We can see that, even in the most distinguishable case, total mass $=3\,M_\odot$, to distinguish the two models using one event observed by aLIGO\_DESIGN, the luminosity distance needs to be $\lesssim \unit{25}{\mega\parsec}$, and even closer for lower masses.}
\label{fig:distance_aLIGO}
\end{figure}

\begin{figure*}
\resizebox{\linewidth}{!}{\includegraphics{"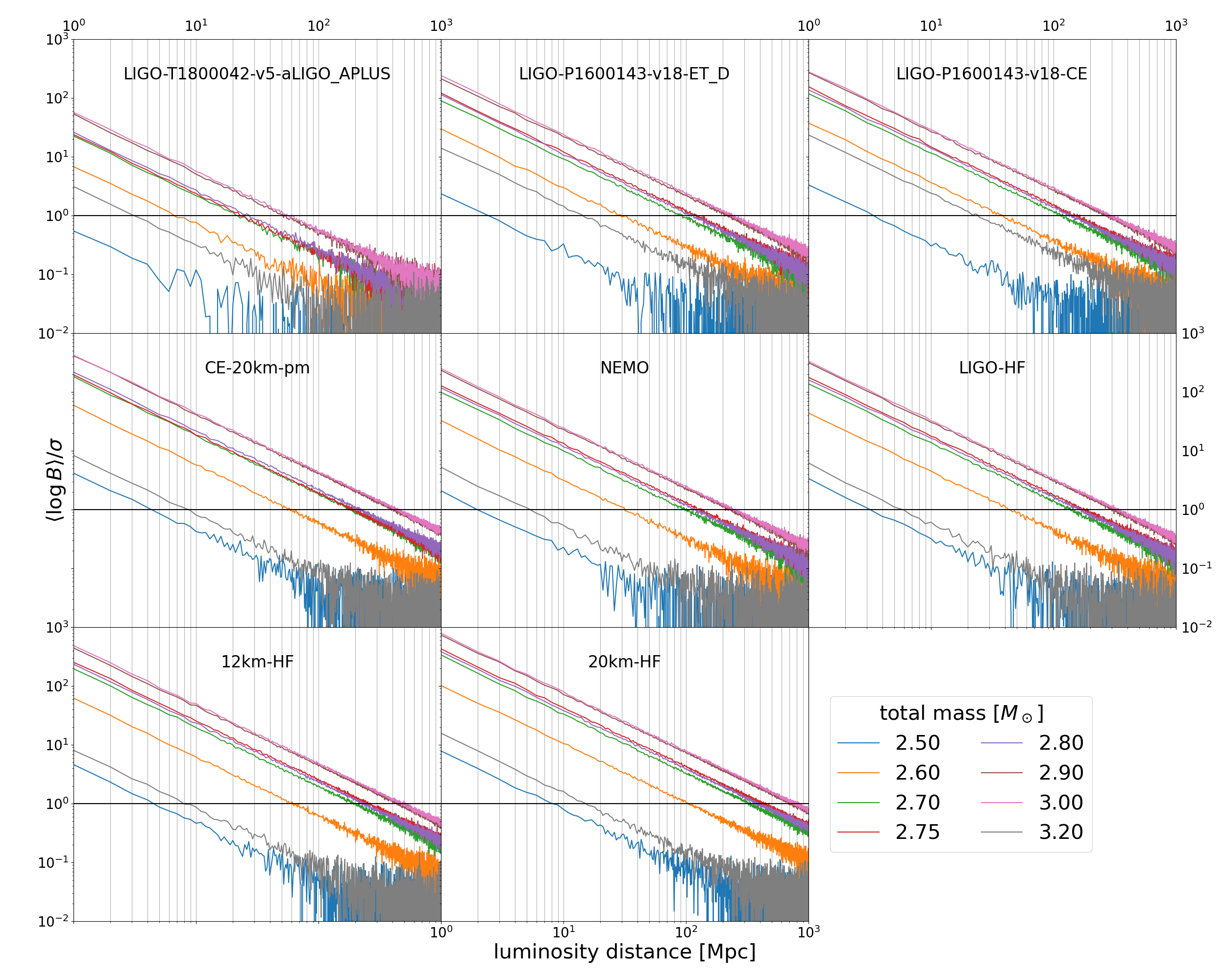"}}
\caption{$D_L$ \textit{vs.}\ $\langle \log B \rangle/\sigma$ for various detectors and total masses assuming optimal orientation for ``+'' polarization.
All binaries plotted here are equal-mass systems.
The legend shows the total mass of the binaries corresponding to each line in units of solar mass.
Third generation detectors and detectors specialized for high frequencies can be expected to distinguish the two scenarios even for events at distances $\sim \unit{100}{\mega\parsec}$.}
\label{fig:distance_totalmass}
\end{figure*}

\begin{figure*}
\resizebox{\linewidth}{!}{\includegraphics{"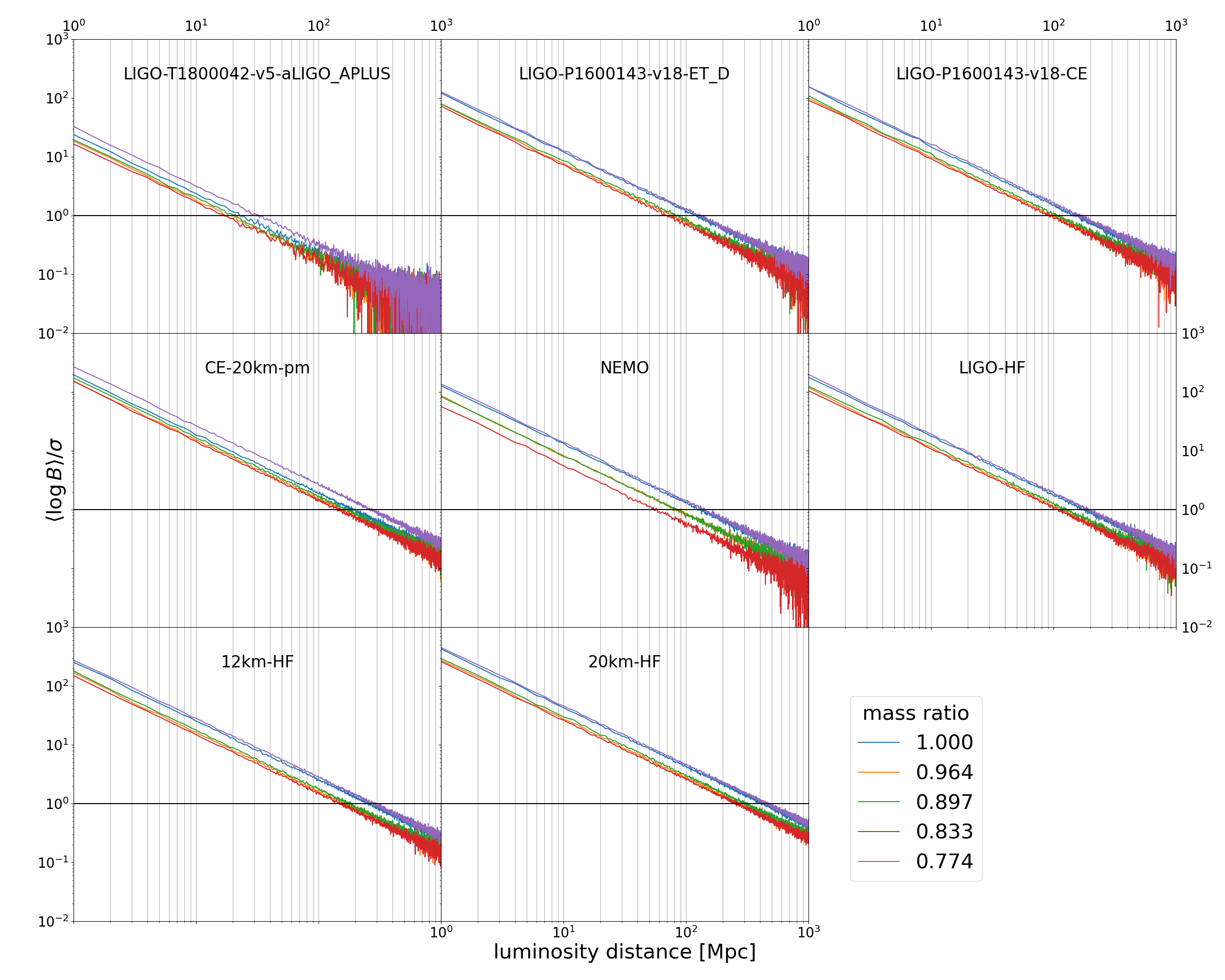"}}
\caption{$D_L$ \textit{vs.}\ $\langle \log B \rangle/\sigma$ plots for various detectors and mass ratios assuming optimal orientation for ``+'' polarization.
For all mass configurations plotted here, the total mass is $2.75\,M_\odot$.
The legend shows the mass ratio $q\equiv m_2/m_1$ of the binaries corresponding to each line.
Although the results vary somewhat due to the change in the mass ratio, it is found to be less influential than the change in the total mass.}
\label{fig:distance_massratio}
\end{figure*}

This test was performed for various detectors and masses and the plots shown in \figref{fig:distance_aLIGO}, \figref{fig:distance_totalmass} and \figref{fig:distance_massratio} are obtained.
The relation of  \eqref{eq:indicator_DL} can also be confirmed from these plots.
As a result, we found that aLIGO\_DESIGN would not help our goal of distinguishing the two quark-hadron transition scenarios, but third generation detectors, NEMO and detectors proposed in \cite{Martynov_2019} are promising.
We discuss these results below.

\figref{fig:distance_aLIGO} and \figref{fig:distance_totalmass} show the results for equal-mass configurations for aLIGO\_DESIGN and the other detectors, respectively.
We find that the case $(m_1,\,m_2)=(1.5,\,1.5)\,M_\odot$ (total mass $=3\,M_\odot$) has the best distinguishability in all panels in \figref{fig:distance_aLIGO} and \figref{fig:distance_totalmass}.
This is related to the fact that, as mentioned in \secref{sec:introduction}, the nature of the quark-hadron transition can be inferred by the time it takes to collapse into a black hole.
The remnants of heavier binaries gravitaionally collapse earlier.
For a remnant of exquisite mass that collapse quickly in the scenario \wCO and not in \woCO, the difference in the stiffness of EoS between these two transition scenarios appears prominently in the gravitational waveform.
In the case of $(1.25,\, 1.25)\,M_\odot$ shown in \figref{fig:waveform}, remnants of less massive binaries do not collapse regardless of the occurrence of crossover so that gravitational waveforms are practically the same for these EoSs.
The same goes for remnants of too massive binaries, such as $(1.6,\, 1.6)\,M_\odot$ shown in \figref{fig:waveform}, because they collapse immediately in either transition scenario.

The line for $(1.5,\,1.5)\,M_\odot$ in the aLIGO\_DESIGN plot (\figref{fig:distance_aLIGO}) intersects the $\langle\log B\rangle/\sigma=1$ line at $D_L\sim\unit{25}{\mega\parsec}$.
In \cite{PhysRevX.13.011048}, the BNS merger rate was found to be $(10\,\text{--}\,1700)\,\mathrm{Gpc^{-3}a^{-1}}$.
Then the expected number of events with $D_L\lesssim \unit{25}{\mega\parsec}$ is $(0.00065\,\text{--}\,0.11)\,\mathrm{a}^{-1}$.
Because the results in this article were obtained assuming a Dirac delta function prior and the optimal orientation, in general this test will be more challenging.

The many lines in the ET\_D plot, \textit{e.g.}, total mass $= 2.70\,M_\odot$, $2.75\,M_\odot$ and $2.80\,M_\odot$, intersect the $\langle\log B\rangle/\sigma=1$ line around $D_L\sim\unit{100}{\mega\parsec}$.
When $(m_1 ,\, m_2 ) = (1.375,\, 1.375)\,M_\odot$ and $D_L = \unit{100}{\mega\parsec}$, SNR of the inspiral phase $\sim412$ as mentioned after  (\ref{eq:Fisher}).
The expected number of events with $D_L \lesssim \unit{100}{\mega\parsec}$ is $(0.042\,\text{--}\,7.1)\,\mathrm{a}^{-1}$.
Thus, the chance of distinguishing the transition scenario is promising for ET\_D, particularly GW170817 has already been observed at $\unit{40}{\mega\parsec}$.

As can be seen from \figref{fig:distance_totalmass}, the high-frequency specialized detectors are as useful as the third generation detector in testing the quark-hadron transition.
When $(m_1 ,\, m_2 ) = (1.375,\, 1.375)\,M_\odot$ and $D_L = \unit{100}{\mega\parsec}$, SNR of the inspiral phase $\sim54$ for NEMO and $\sim71$ for LIGO-HF.
Since these detectors are not as sensitive at low frequencies as ET\_D, the SNR of this kind of signals for these detector are not so great but still higher than that of GW170817.
One can also expect the parameters to be well constrained by networks composed of other detectors.

\subsubsection{Dependence on Mass Ratio}
We also investigated how the distinguishability behaves when the mass ratio varies.
The five lines in panels of \figref{fig:distance_massratio} correspond to a total mass of 2.75 $M_\odot$ and five different mass ratios.
Although the results vary somewhat due to the change in the mass ratio, it is found to be less influential than the change in the total mass.
This is a positive result because the total mass can usually be measured more accurately than the mass ratio, as is the case with the analysis results for GW170817 \cite{GW170817_observation}.

\subsection{Distinguishability \textit{vs.}\ Event Rate}
\begin{figure*}
\resizebox{\linewidth}{!}{\includegraphics{"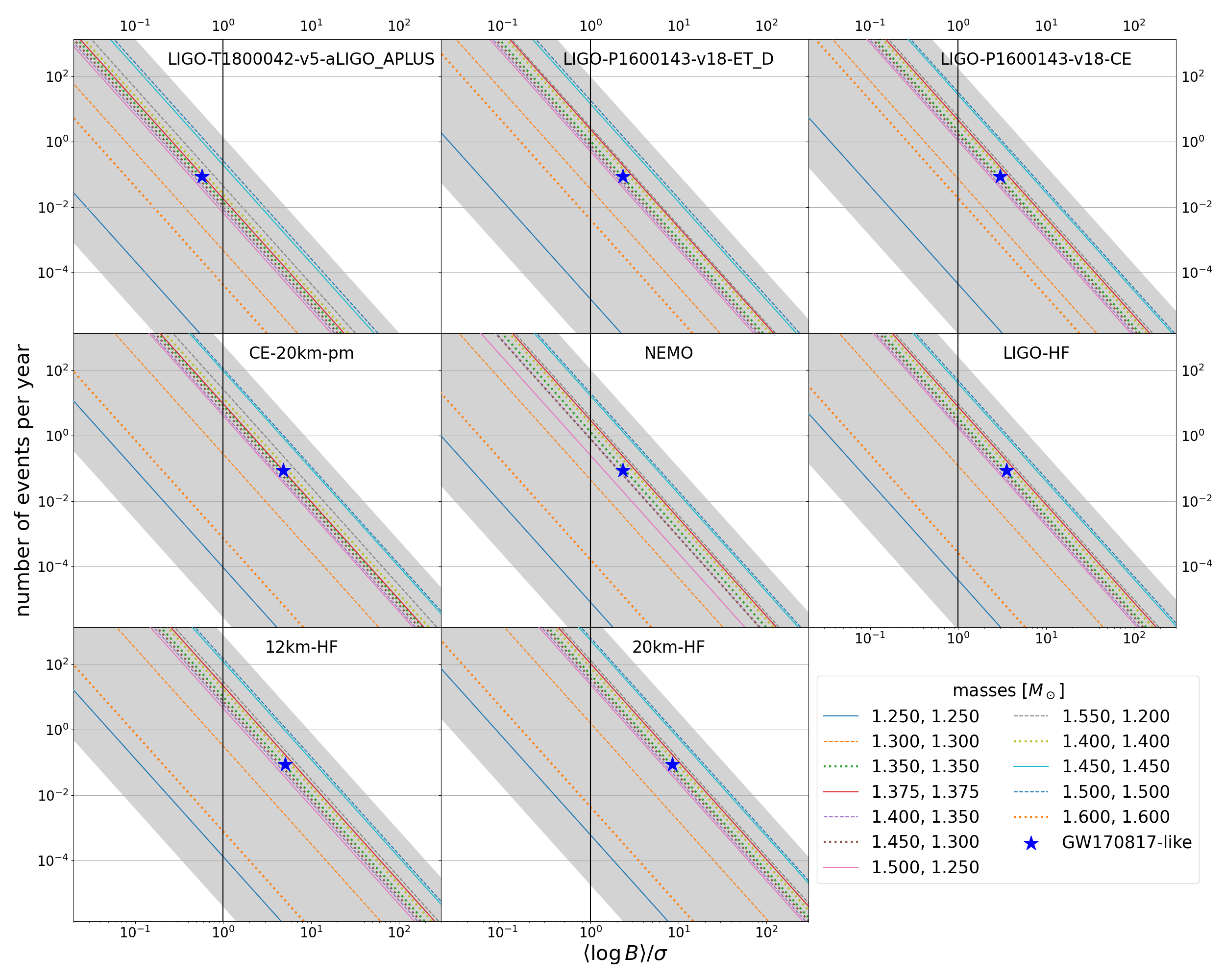"}}
\caption{$\langle\log B\rangle/\sigma$ \textit{vs.}\ event rate plots.
The colored lines correspond to each mass configuration and the merger rate of $330\,\mathrm{Gpc^{-3}a^{-1}}$.
The bottom edge of the gray-shaded area corresponds to mass configuration $(1.25, \,1.25)\,M_\odot$ and the merger rate $10\,\mathrm{Gpc^{-3}a^{-1}}$.
The upper edge corresponds to mass configuration $(1.5, \,1.5)\,M_\odot$ and the merger rate $1700\,\mathrm{Gpc^{-3}a^{-1}}$.
Stars represent events consistent with the estimated mass and luminosity distance of GW170817.
The legend shows the components mass of the binaries corresponding to each line in units of solar mass.}
\label{fig:rate}
\end{figure*}
For visibility, we converted distance \textit{vs.}\ $\langle \log B\rangle/\sigma $ plots in \figref{fig:distance_totalmass} and \figref{fig:distance_massratio} to $\langle \log B \rangle/\sigma$ \textit{vs.}\ event rate plots in \figref{fig:rate}.
To do this, we performed least-squares fittings as in \figref{fig:distance_155_12_ET_D}.
The colored lines in \figref{fig:rate} correspond to each mass configuration and the merger rate of $330\,\mathrm{Gpc^{-3}a^{-1}}$, which is the geometric mean of the range of the merger rate $(10\,\text{--}\,1700)\,\mathrm{Gpc^{-3}a^{-1}}$.
The bottom edge of the gray-shaded area corresponds to mass configuration $(1.25, \,1.25)\,M_\odot$ and the lower limit of merger rate $10\,\mathrm{Gpc^{-3}a^{-1}}$.
The upper edge corresponds to mass configuration $(1.5, \,1.5)\,M_\odot$ and the upper limit of merger rate $1700\,\mathrm{Gpc^{-3}a^{-1}}$.
For reference, the points consistent with GW170817's estimated distance $D_L = \unit{40}{\mega\parsec}$, chirp mass $M_c=1.188^{+0.004}_{-0.002}\,M_\odot$, and mass ratio $m_2/m_1=(0.4\,\text{--}\,1.0)$ (90\% credible intervals) \cite{GW170817_observation} are plotted in \figref{fig:rate} as stars.
To be exact, the stars plotted at the event rate of
\begin{equation}
	330\times \frac{4\pi}{3}\left(\frac{40}{1000}\right)^3=0.088\,\mathrm{a^{-1}}
\end{equation}
and the value of $\langle \log B \rangle/\sigma$ of 
\begin{equation}
	\frac{
		\left.\frac{\langle \log B \rangle}{\sigma}\right|_{1.45,\,1.3} +
		\left.\frac{\langle \log B \rangle}{\sigma}\right|_{1.5,\,1.25} +
		\left.\frac{\langle \log B \rangle}{\sigma}\right|_{1.55,\,1.2}
	}{3},
	\label{eq:average}
\end{equation}
where 
\begin{equation}
	\left.\frac{\langle \log B \rangle}{\sigma}\right|_{a,\,b}\equiv
	\left.\frac{\langle \log B \rangle}{\sigma}\right|
	_{D_L=\unit{40}{\mega\parsec},\,(m_1,\,m_2)=(a,\,b)\,M_\odot}.
\end{equation}
These masses in  (\ref{eq:average}) were chosen because they are relatively close to the 90\% credible intervals of GW170817 above among the available NR waveforms.

The gray shading in panels of \figref{fig:rate} for the third generation detectors and the detectors specialized for high frequencies covers the region of $\langle\log B \rangle/\sigma\geq 1$ and event rates tens to hundreds per year.
It was found that there is a realistic possibility to judge the quark-hadron transition scenario using these detectors.


\subsection{Results of Dimensionality Reduction}\label{subsec:rodr}
\begin{figure*}
\resizebox{.9\linewidth}{!}{\includegraphics{"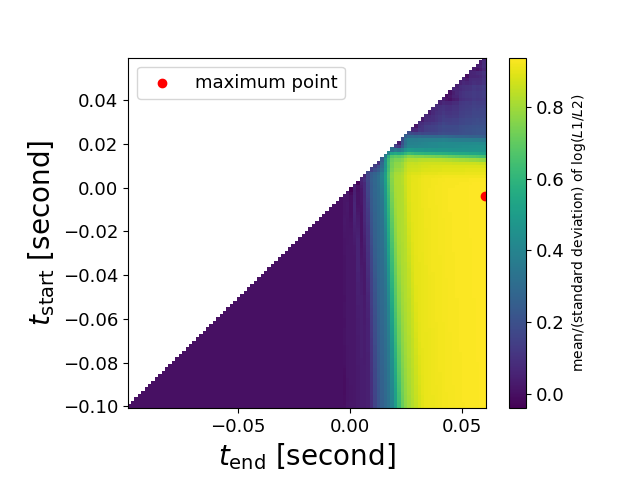"}%
\includegraphics{"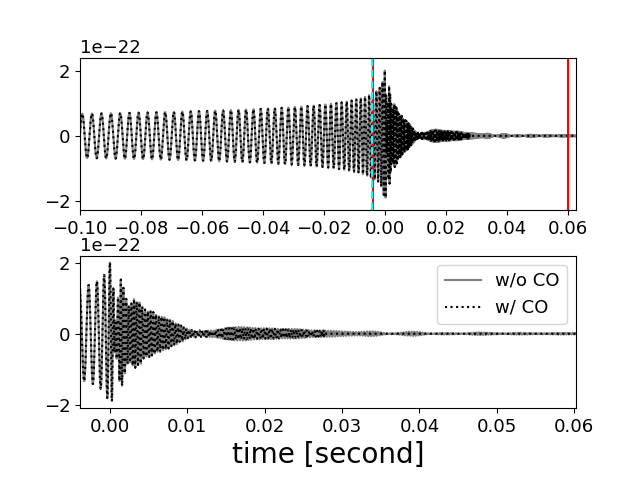"}}
\caption{The results of search for optimal interval in the case of $(m_1,\,m_2)=(1.35,\,1.35)\,M_\odot,\,D_L=\unit{100}{\mega\parsec}$, the optimal orientation for ``+'' polarization and ET\_D.
The red point on the heatmap shows the interval $[t_\mathrm{start}, t_\mathrm{end}]$ giving the maximum value of $\langle \log B\rangle/\sigma$.
This interval is indicated by two red lines in the upper right panel, and shown in more detail in the lower right panel.
The dashed cyan line in the upper right panel indicates the connection of the analytic and the NR waveform.
}
\label{fig:egheatmap}
\end{figure*}
\begin{figure*}
	\resizebox{0.8\linewidth}{!}{\includegraphics{"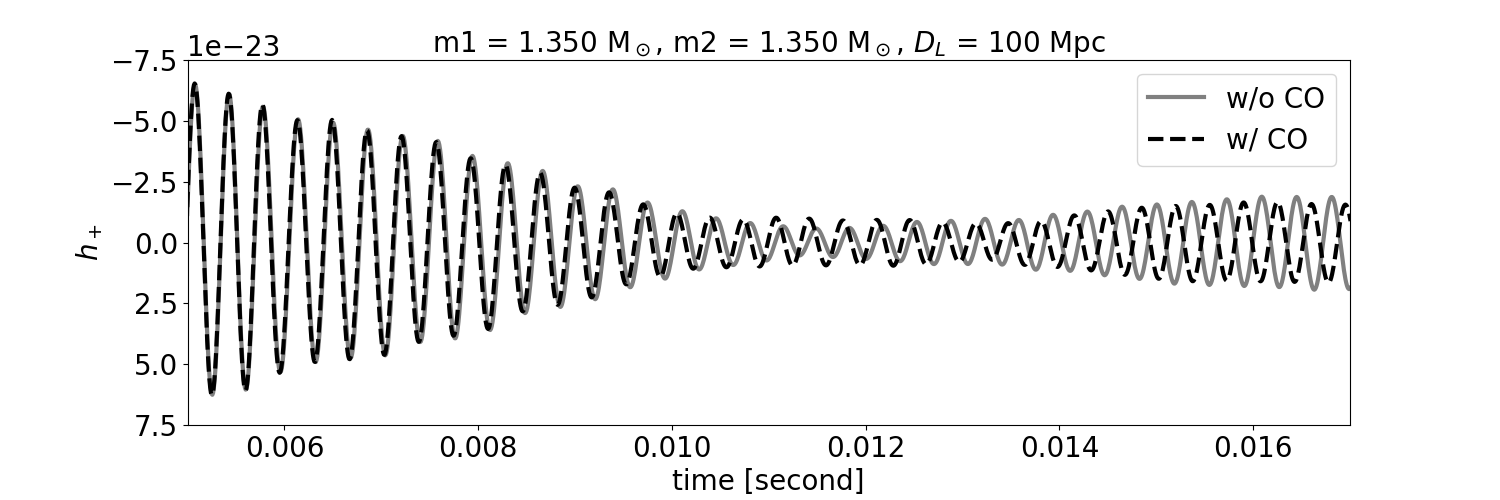"}}
	\caption{Detailed view of waveform around $t=\unit{0.01}{\second}$. $(m_1,\,m_2)=(1.35,\,1.35)\,M_\odot$ and $D_L=\unit{100}{\mega\parsec}$.} 
	\label{fig:detailedwf}
\end{figure*}
\begin{figure*}
	\resizebox{.45\linewidth}{!}{\includegraphics{"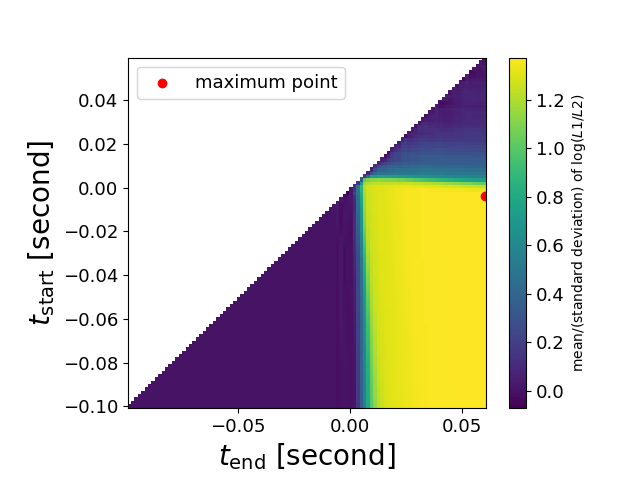"}}
	\resizebox{.45\linewidth}{!}{\includegraphics{"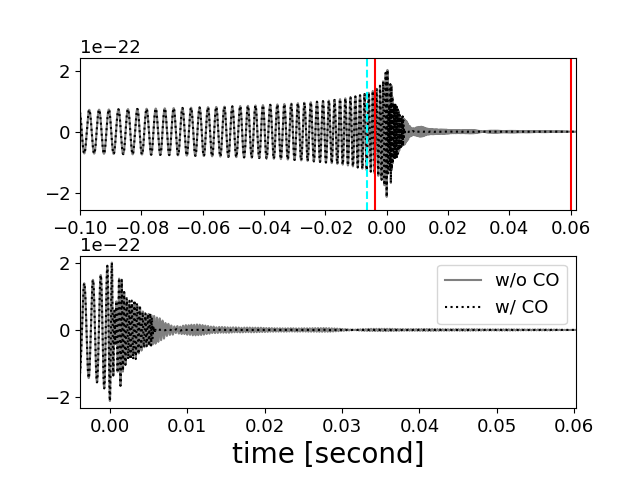"}}
	\caption{The results of search for optimal interval in the case of $(m_1,\,m_2)=(1.4,\,1.4)\,M_\odot,\,D_L=\unit{100}{\mega\parsec}$, the optimal orientation for ``+'' polarization and LIGO-P1600143-v18-CE.
	}
	\label{fig:egheatmap_CE}
\end{figure*}
\begin{table*}
	\caption{Improvements in $\langle \log B\rangle/\sigma$.  Each cell in this table shows the change in the value of
	$\langle \log B\rangle/\sigma$ from $[t_\mathrm{start},\,t_\mathrm{end}]=[-0.1\,\mathrm{s},$ the end of NR waveform$]$ to
	$[t_\mathrm{start},\,t_\mathrm{end}]=$ maximum point.
	The injected waveforms are simulated as emitted at a luminosity distance of $\unit{100}{\mega\parsec}$, and the corresponding masses are listed in the leftmost column.
	CE here is LIGO-P1600143-v18-CE (\unit{40}{\kilo\meter} configuration).}
	\label{table:logB1}
	\begin{center}
		\begin{tabular}{|c||c|c|c|c|c|c|c|c|}
			\hline
			$m_1,\,m_2\,[M_\odot]$ & aLIGO & A+ &ET\_D& CE& NEMO&LIGO-HF&12km-HF&20km-HF\\
			 \hline\hline
			1.25, 1.25 & $0.005\rightarrow0.07$& $0.004\rightarrow0.08$& $0.01\rightarrow0.09$& $0.05\rightarrow0.1$& $-0.04\rightarrow0.06$& $0.03\rightarrow0.09$& $0.07\rightarrow0.1$& $0.09\rightarrow0.1$\\
			1.3, 1.3 & $0.01\rightarrow0.09$& $0.05\rightarrow0.07$& $0.31\rightarrow0.31$& $0.40\rightarrow0.40$& $0.27\rightarrow0.34$& $0.44\rightarrow0.45$& $0.64\rightarrow0.64$& $1.0\rightarrow1.0$\\
			1.35, 1.35& $0.08\rightarrow0.1$&$0.3\rightarrow0.3$& $0.93\rightarrow0.94$& $1.1\rightarrow1.1$& $1.0\rightarrow1.1$& $1.4\rightarrow1.4$& $2.0\rightarrow2.0$& $3.2\rightarrow3.2$\\
			1.375, 1.375& $0.1\rightarrow0.1$&$0.2\rightarrow0.2$& $1.0\rightarrow1.0$& $1.2\rightarrow1.2$& $1.0\rightarrow1.1$& $1.4\rightarrow1.4$& $2.1\rightarrow2.1$& $3.4\rightarrow3.4$\\
			1.4, 1.35& $0.06\rightarrow0.1$&$0.1\rightarrow0.2$& $0.75\rightarrow0.75$& $0.95\rightarrow0.95$&$0.81\rightarrow0.82$& $1.1\rightarrow 1.1$& $1.7\rightarrow1.7$& $2.8\rightarrow2.8$\\
			1.45, 1.3& $0.1\rightarrow0.1$&$0.2\rightarrow0.2$& $0.86\rightarrow0.87$& $1.1\rightarrow1.1$&$0.75\rightarrow0.88$& $1.3\rightarrow 1.3$& $1.7\rightarrow1.7$& $3.0\rightarrow3.0$\\
			1.5, 1.25& $0.05\rightarrow0.1$&$0.2\rightarrow0.2$& $0.79\rightarrow0.79$& $0.92\rightarrow0.92$&$0.56\rightarrow0.80$& $1.2\rightarrow 1.2$& $1.5\rightarrow1.5$& $2.6\rightarrow2.6$\\
			1.55, 1.2& $0.1\rightarrow0.1$& $0.4\rightarrow0.4$& $1.4\rightarrow1.4$& $1.7\rightarrow1.7$& $1.4\rightarrow1.4$&$1.9\rightarrow1.9$ &$ 2.8\rightarrow2.8$& $4.8\rightarrow4.8$\\
			1.4, 1.4& $0.1\rightarrow0.1$&$0.3\rightarrow0.3$& $1.1\rightarrow1.1$& $1.4\rightarrow1.4$&$1.1\rightarrow1.2$& $1.6\rightarrow1.6$& $2.3\rightarrow2.3$& $3.8\rightarrow3.9$\\
			1.45, 1.45& $0.2\rightarrow0.2$&$0.5\rightarrow0.6$& $2.3\rightarrow2.3$& $2.7\rightarrow2.7$&$2.2\rightarrow2.2$& $3.0\rightarrow3.0$& $4.4\rightarrow4.4$& $7.4\rightarrow7.4$\\
			1.5, 1.5& $0.3\rightarrow0.3$&$0.6\rightarrow0.6$& $2.5\rightarrow2.5$& $3.0\rightarrow3.0$& $2.2\rightarrow2.4$& $3.2\rightarrow3.2$& $4.7\rightarrow4.7$& $7.8\rightarrow7.9$\\
			1.6, 1.6& $0.02\rightarrow0.04$&$0.06\rightarrow0.08$& $0.1\rightarrow 0.1$ & $0.2\rightarrow0.2$& $0.04\rightarrow 0.06$& $0.07\rightarrow0.09$& $0.09\rightarrow0.1$& $0.1\rightarrow0.2$\\
			\hline
		\end{tabular}
	\end{center}
\end{table*}

As described in \secref{subsec:dimensionmethod}, we computed $\langle \log B \rangle/\sigma$ for various time intervals $[t_\mathrm{start} , t_\mathrm{end}]$ and examined its behavior.
Here we summarize the results.

Figure \ref{fig:egheatmap} shows the results in the case of $(m_1,\,m_2)=(1.35,\,1.35)\,M_\odot,\,D_L=\unit{100}{\mega\parsec}$, the optimal orientation for ``+'' polarization and ET\_D.
The red point on the heatmap shows the interval $[t_\mathrm{start}, t_\mathrm{end}]$ giving the maximum value of $\langle \log B\rangle/\sigma$.
The interval between the two red straight lines in the upper right panel and the interval shown in the lower right panel correspond to the interval represented by the red points on the heatmap.
However, no significant difference in $\langle \log B\rangle/\sigma$ values was observed between this maximum point and the lower right corner of the heatmap corresponding to the longest time interval.
A characteristic feature of the heatmap in \figref{fig:egheatmap} is that $\langle \log B\rangle/\sigma\sim0$ for data not including $t\gtrsim\unit{0.01}{\second}$, and conversely, the $\langle \log B\rangle/\sigma$ value is also 0 for data including only $t\gtrsim\unit{0.01}{\second}$.
As \figref{fig:detailedwf} and the right panel of \figref{fig:egheatmap} show, around $t\sim\unit{0.01}{\second}$, the amplitudes become small and the difference between the phases of  two models begins to be prominent because of the onset of crossover.
Then, it can be interpreted that the former is due to the small difference between the two models, and the latter is due to the small SNR.
The same tests were performed with other detector/mass combinations and yielded heatmaps of similar structures (for example, \figref{fig:egheatmap_CE}).
None of the combinations showed significant improvement in the $\langle \log B\rangle/\sigma$ value by reducing the dimensionality of the data as shown in TABLE \ref{table:logB1}.

\section{Conclusion}\label{sec:conclusion}

In this study, we investigated whether some of the proposed gravitational wave detectors can distinguish two models constructed in the previous study \cite{PhysRevLett.130.091404}, a  model that smoothly connects the EoS of a nuclear branch predicted by $\chi$EFT \cite{Drischler_2021} and a quark branch predicted by pQCD (\wCO), and a model without such a crossover (\woCO).
As a result, it was found that to distinguish the two models using one event observed by aLIGO\_DESIGN \cite{LIGO-T1800044}, an event within a luminosity distance of $\unit{25}{\mega\parsec}$ is required, and such an event is thought to occur only once every several decades \cite{PhysRevX.13.011048}.
However, the third generation detectors \cite{Hild_2011,LIGO-P1600143, Abbott_2017, Srivastava_2022} and the high frequency specialized detectors, such as NEMO \cite{Ackley_2020}, LIGO-HF, 12km-HF and 20km-HF \cite{Martynov_2019} are promising.
These detectors are expected to observe a few events per year that can distinguish the two models if parameters such as mass are well determined by the analysis of the inspiral phase.
Furthermore, a rare but close event will be quite useful for high-credibility distinction of the scenarios.

We also investigated whether the dimensionality reduction of the data can improve the distinguishability of two candidates.
We found that the distinguishability, which is evaluated based on the value of $\langle \log B \rangle/\sigma$ (see \secref{subsec:criterion}), hardly depends on the length of the pre-merger waveforms used in the calculation.
Although the original objective of improving the distinguishability of the models by reducing the dimension was not met, this fact also implicates that the results we got are robust.
Furthermore we were able to get some insight into where is the  useful part of waveform for this comparison.

All the results of this study are based on the assumption that the equation of state of hadronic matter close to the saturation density is well determined and that parameters such as mass can be determined with high precision by analysis of gravitational waves in the inspiral stage.
A future task is to conduct tests in a more realistic setting.

\acknowledgments
The authors are grateful for computational resources provided by the LIGO Laboratory and supported by National Science Foundation grants PHY-0757058 and PHY-0823459.
This research has made use of LALSuite software \cite{lalsuite}.
This work was supported by JST SPRING, grant number JPMJSP2108, JST FOREST, grant number JPMJFR2136, Japan Society for the Promotion of Science (JSPS) Grants-in-Aid for Scientific Research (KAKENHI) grant numbers JP17H06361, JP18H03698, JP20H00158, JP20H05639, JP22K03617 and JP23H04900.

\bibliography{article}
\end{document}